\documentclass[12pt,preprint]{aastex63}
\begin{document}

\title{ROME (Radio Observations of Magnetized Exoplanets). II. HD 189733 Does Not Accrete Significant Material from Its Exoplanet Like a T Tauri Star from A Disk}

\author[0000-0001-6987-6527]{Matthew Route}
\affiliation{Research Computing, Purdue University, 155 S. Grant St., West Lafayette, IN 47907, USA}

\author[0000-0002-4540-6587]{Leslie W. Looney}
\affiliation{Department of Astronomy, University of Illinois, 1002 West Green Street, Urbana, IL 61801, USA}

\correspondingauthor{Matthew Route}
\email{mroute@purdue.edu}


\received{2019 September 30}
\revised{2019 November 18}
\accepted{2019 November 18}
\published{2019 December 23}
\submitjournal{The Astrophysical Journal}
\reportnum{ApJ 2019, 887, 229}

\begin{abstract}
It has been asserted that the primary star in the HD 189733 system steadily accretes evaporated exospheric gases from its ``hot Jupiter'' companion, rather like a T Tauri star accreting from a disk.  We conduct statistical and periodogram analyses of the photometric time series of the primary, as acquired by the automated photoelectric telescope (APT), \emph{Microvariability and Oscillations of Stars} (\emph{MOST}), and Wise Observatory, to investigate this claim with the goal of revealing the presence of accretion shocks or photospheric accretion hotspots as are found in T Tauri systems such as AA Tau.  None of the anticipated features were found.  We re-analyze existing radio, optical, ultraviolet, and X-ray data within the framework of accreting T Tauri systems to determine physical quantities such as plasma density and temperature, accretion rate, and flare lengths.  We find that with an accretion rate of $\dot{M}\sim10^{9}$ to 10$^{11}$ g s$^{-1}$, the star is more similar to a system that intermittently absorbs gas from sungrazing comets in outburst than classical T Tauri systems, which have accretion rates at least two orders of magnitude larger.  If such accretion exists, it would result in undetectably low activity at all wavelengths.  Alternatively, all of the emission properties observed thus far are in agreement with stellar activity from a magnetically active star.
\end{abstract}

\section{Introduction}
As interest in ``star-planet interactions'' has grown, attention has focused on the HD 189733 system, which has been the subject of many observing campaigns across multiple wavelengths.  The system consists of three components: a K1.5V primary star ($M_{\ast}$=0.82$M_{\odot}$, $R_{\ast}$=0.76$R_{\odot}$) located $d$=19.25 pc away, a $\sim$M4V secondary star separated from the primary by $\sim$220 au, and a 1.13$\pm$0.03 $M_{J}$ exoplanet that orbits the primary every 2.2185733$\pm$0.0000019 days at a distance of 0.031 AU (8.77 $R_{\ast}$) \citep{bou05,bak06,win07,boi09}.  The primary has chromospheric activity index $S$-value=0.525, making it very active among K dwarfs \citep{wri04}.

``Star-planet interactions'' is a term used to describe how exoplanets that orbit close to their host stars may enhance activity on those stars via tidal and magnetic interactions.  The flip side of this interaction is that stellar irradiation, stellar winds, stellar magnetic activity cycles, and stellar evolution in general affect the development and evolution of the exoplanets themselves, with consequences for the climate of potentially habitable worlds, and possibly, the development and evolution of life on them (e.g., \cite{edd76,kal17,lea18,woo18,jon19}).  The first suspected systems to host ``star planet interactions'' were those with Jupiter-mass exoplanets in several-day orbits, called ``hot Jupiters.''  The possibility that several ``hot Jupiters'' may cause periodic chromospheric and coronal emissions from their host stars has been a subject of debate (see \citet{fra18} for a comprehensive review).  Among these, the HD 189733A/b system has been particularly well-studied, with observations conducted at radio, optical, ultraviolet (UV), and X-ray wavelengths, as thoroughly recounted and examined in our earlier work, ROME I \citep{rou19}.  ROME I was the first paper in our ROME series, which presents and examines Radio Observations of Magnetized Exoplanets in the context of ``star-planet interactions.''

In ROME I, we critically examined several hypotheses regarding the stellar activity of HD 189733A.  These hypotheses included that HD 189733A exhibits enhanced chromospheric activity in \ion{Ca}{2} H\&K at orbital phase $\phi_{orbit}\sim$0.8, that HD 189733A displays enhanced coronal flaring in X-rays at $\phi_{orbit}\sim$0.52-0.65, and that persistent starspots exist in the photosphere.  Our analysis of stellar activity on the primary mainly consisted of presenting sensitive, high-temporal resolution 4.2-5.2 GHz radio observations conducted at Arecibo Observatory and photometric data from the automated photoelectric telescope (APT), \emph{Microvariability and Oscillations of Stars} (\emph{MOST}) satellite, and Wise Observatory. The radio data spanned orbital phases $\phi_{orbit}$=0.568-0.608 and failed to detect any modestly circularly polarized flaring emission that could be the counterpart of X-ray flares.  Our photometric analysis investigated whether there were temporal trends among the highly magnetized active regions, which would be manifested at optical wavelengths as starspot groups, and therefore cause darker than average regions on the stellar surface.  However, this analysis indicated that the orbital phase range alleged to contain enhanced coronal activity was not any darker than average, and therefore did not exhibit enhanced starspot formation or persistent magnetic activity above the average background rate.  Furthermore, by combining these data sets with a reexamination of archival data at other wavelengths, we developed a comprehensive picture of stellar activity within the system.  Using arguments based on physical processes, multiwavelength phenomenology, and statistical analysis, we found that there was no compelling case for these three hypotheses, and in each case, we found significant reasons to doubt them.

In ROME II, we return to the HD 189733 system to evaluate the most provocative assertion about the system, namely that \emph{``a stream of gas evaporating from the planet is actively and almost steadily accreting onto the stellar surface, impacting at 70$\degr$-90$\degr$ ahead of the subplanetary point''} \citep{pil15}.  This assertion was buttressed by several lines of argument.  First, X-ray flares were observed at several epochs in a restricted orbital phase range: $\phi_{orbit}\sim$0.54, 0.52, and 0.64 \citep{pil10,pil11,pil14}.  Second, \citet{pil11,pil14} suggested that the corona of HD 189733A is cooler than the coronae of other main sequence stars, yet is denser and more luminous as inferred from \ion{O}{7} and \ion{Ne}{9} spectral lines.  The corona may therefore be intermediate between that of a main sequence star and that of an accreting pre-main sequence (PMS), or classical T Tauri star.  Third, \cite{pil14} computed flare properties and oscillation timescales that implied flares stretching $\sim$25\% of the distance between the star and exoplanet.  Fourth, \emph{Hubble Space Telescope (HST)} Cosmic Origins Spectrograph (COS) data at far ultraviolet (FUV) wavelengths observed the presence of two episodes of flaring at $\phi_{orbit}\sim$0.525 and $\phi_{orbit}\sim$0.588 that appear to coincide with the activity observed in X-rays.  From these events, \citet{pil15} hypothesized that the evaporating exosphere of HD 189733b forms an accretion stream of $\sim2 R_{\ast}$ in length.  This accretion stream falls supersonically toward the primary star, where an outgoing stellar wind causes a shock, leading to a bent ``knee'' formation in the accretion stream (see their Figure 12).  The shocked material is then dragged to the stellar photosphere by gravity, where it creates a small hotspot.  \citet{pil10,pil14} noted that the activity of HD 189733A implies a young star system, while the X-ray luminosity of its M4 dwarf companion implies an older system.  This could indicate the transmutation of exoplanetary orbital angular momentum into primary star rotational angular momentum via tidal interactions.  Alternatively, the presence of the ``hot Jupiter'' magnetosphere may lead to the closure of otherwise open magnetic field lines, thereby inhibiting primary star spindown \citep{lan10}.

A key aspect of these assertions is the proposed similarities between the HD 189733 system and T Tauri systems.  T Tauri star systems can be divided into two classes: classical T Tauri stars (CTTSs) that are actively accreting and weak T Tauri stars (WTTSs) that are dominated by stellar magnetic activity.  Initially, these two classes were divided by a simple observational distinction: CTTSs have an H$\alpha$ equivalent width $W_{eq}$(H$\alpha$)$>$10\AA, while WTTS have the complementary property \citep{her88}.  An alternative definition is based on the slope of their spectral energy distributions from 2.2-25 $\mu$m, $\alpha_{IR}=d log(\lambda F_{\lambda})/d log(\lambda)$, where CTTSs have -1.6$<\alpha_{IR}<$-0.3 and WTTSs have $\alpha_{IR}<$-1.6 \citep{dun14}.  These properties indicate that CTTSs are actively accreting pre-main-sequence stars with significant disks, many of which exhibit inverse P-Cygni profiles, while WTTSs are weakly or non-accreting.

\citet{pil11} compared the \ion{O}{8}/\ion{O}{7} line ratios from HD 189733 to those found in main sequence (MS) and CTTSs.  They noted that CTTS accretion shocks cause cool X-ray emission.  They found that HD 189733A \emph{``is marginally cooler than MS stars of comparable luminosity, but its corona is overdense like PMS accreting stars.''}  From their measurements and analysis they concluded that the corona of HD 189733A is more active and luminous than the solar corona, which could be caused by ``star-planet interactions'' between it and its ``hot Jupiter'' companion.  Similarly, \citet{pil14} found an anomalously overdense corona and supported their description with their Figure 6 that compares HD 189733A with MS stars and CTTSs.  They also found that its X-ray variability is stronger than that found in field stars, but closely resembles that observed on young active stars.  Finally, \citet{pil15} presented a detailed scenario featuring accretion stream structures that fit their FUV and X-ray data.  In particular, they noted that \emph{``In HD 189733 we observe narrow lines like in WTTSs and a systematic redshift of C II lines that is usually associated with emission from the gas flowing down in the coronal loops.''}

In this paper, we will examine our photometric and radio data within the context of the T Tauri paradigm suggested by \citet{pil11,pil14,pil15}.  In particular, given the scenario presented by \citet{pil15} above, there are several reasons to suspect that HD 189733A may have stable, bright regions phased to the orbital velocity of the exoplanet.  First, we note that their model describes the presence of two potential hotspots: the accretion shock, or ``knee'' feature, and the photospheric hot spot where the shocked gas impacts the photosphere.  Both features may be visible at optical wavelengths and revealed by a reanalysis of existing photometric data.  

Second, we note that the literature on T Tauri stars provides examples of photometric variability related to accretion flows.  Photometric observing campaigns of the CTTS SU Aur found large, sudden decreases in mean light levels of the range $\Delta V\sim$0.8 mag.  This variability was attributed to the $P\sim$1.7 d spot-induced rotation modulation of the star accompanied by random episodes of dimming caused by concentrations of dusty matter in a non-axisymmetric, relatively cool, accretion disk \citep{dew03}.  The three observing campaigns of the CTTS AA Tau (see \citet{bou07} and references therein) comprise the most thorough extant analysis of T Tauri star photometry.  From these observations, \citet{bou07} were able to disentangle the effects of the 8.2 d spot-modulated rotation of the star, the periodic occultation of the photosphere by a warped, inner dust disk that borders on the corotation radius, and features associated with accretion funnel flows and shocks.  Their observations indicated that an accretion flow would manifest as a broad photometric minimum of variable depth, such that greater accretion rates correspond to deeper minima.  These features are magnetically controlled and hence, rotate with the stellar rotation period.  Similarly, reconstructions of the large-scale magnetic topology of the CTTSs BP Tau, V2129 Oph, and TW Hya based on Zeeman Doppler Imaging (ZDI) measurements revealed the presence of large, cool active regions of high magnetic field intensity located at the foot of their accretion flows \citep{don08,don11a,don11b}.  

A description of the photometric data that we will use in our analysis follows in Section 2.  Section 3 presents our statistical analyses of the bright region photometry to search for signs of accretion.  In Section 4, we fuse photometric, radio, optical and UV spectral line, and X-ray data to examine the accretion scenario in detail and compare the emission of HD 189733A at these wavelengths to CTTS emissions.  We then summarize the key points of our analysis and propose future testable hypotheses for the HD 189733A accretion scenario in Section 6.  We note that the final point regarding the youth of the HD 189733 system in the above list of supporting evidence for the accretion scenario concerns the dynamics of the system and cannot be assessed with the observations and analysis presented in this paper.  Although an examination of the dynamical history of this system is an intriguing question, we will reserve that analysis for a later paper within the ROME series.

\section{Observations}
To evaluate the claims of steady accretion, we compiled photometric data of the HD 189733 system from four sources: \citet{hen08, mill08, boi09, sing11}.  Data from \citet{hen08} were acquired by the T10 0.8 m APT at Fairborn Observatory, Arizona between 2005 October and 2007 July.  Simultaneous Str\"{o}mgren $b$ and $y$ photometry of the target star and three nearby reference stars were made nightly with 20 s integration times.  From these filters, $V=(b+y)/2$ magnitudes were computed and graphed to display photometric variability.  \citet{mill08} collected photometry from the \emph{MOST} satellite in 2006 August.  The photometry was obtained in Direct Imaging mode through a single broadband filter (3500-7000\AA) with 21 s integrations.  \citet{boi09} duplicated this setup to collect \emph{MOST} photometry from 2007 July to 2007 August, except their time series data were binned in 101.43 min increments.  Finally, \citet{sing11} collected APT and Wise Observatory photometry from 2009 October to 2009 December and 2010 May to 2010 June.  The APT measurements duplicated the setup used in \citet{hen08}, while the Wise Observatory (Israel) photometry were collected through the 1.0 m telescope with an $R$ filter.

From each data set, local maxima that correspond to photometric bright regions were determined by eye and the times that correspond to the midpoints of the local maxima measured.  These measurements were then converted to the rotational cycles of HD 189733A ($E_{Rot}$) and the orbital cycles of exoplanet HD 189733b ($E_{Rot}$) using the ephemerides provided in \citet{far17}:
\begin{equation} T_{0} = HJD~2453629.389 + 12~E_{Rot},\end{equation}
\begin{equation} T_{0} = HJD~2453629.389 + 2.218575~E_{Orb},\end{equation}
These measurements and calculations are reported in Table 1 and graphed in Figure 1.

\section{Results}
A cursory examination of the orbital phases associated with bright regions on HD 189733A reveals no obvious pattern. However, the rotational phases of the bright regions tend to cluster in Figure 1 and persist over several rotational periods. These findings add support to the results found in ROME I that the dark regions, which denote hemispheres with greater starspot activity than average, are correlated over rotational cycles but not orbital cycles.

\subsection{One-Sample Kolmogorov-Smirnov Test}
The 48 bright region timings listed in Table 1 can also be examined in a statistical fashion.  Similar to the analysis used in ROME I, the observed distribution of bright regions can be compared to a model where bright regions are randomly distributed in orbital phase.  We performed a one-sample Kolmogorov-Smirnov test to compare these two distributions of photometric maxima, where this latter continuous uniform distribution constitutes our null hypothesis.  Figure 2 compares these two cumulative distribution functions which yield a $D_{KS}$=0.08, as compared to a critical value of $CV_{KS}$=0.19.  Since we obtain $p$-value=0.91, we fail to reject the null hypothesis that the underlying distributions are the same with $>$0.95 confidence.  It therefore appears that the bright regions exist independently of exoplanet orbital position.

We note from Figure 2 that there appears to be a slight overdensity of bright regions near $\phi_{orbit}\sim$0.3 and a slight dearth of such regions near $\phi_{orbit}\sim$0.8.  These prospective features may be compared with stellar activity observed at radio, optical, UV, and X-ray wavelengths at additional epochs and reported in the literature (see ROME I, Table 2).  In ROME I, Table 4 we determined that the flaring and spotting activity in the $\phi_{orbit}\sim$0.3-0.5 and $\phi_{orbit}\sim$0.7-0.9 orbital phase ranges did not have $>$1.4$\sigma$ statistical significance when compared with a Poisson process that occurs at a definite average rate.  Thus, in ROME I we failed to find a concentration of strongly magnetized, and potentially persistent, active regions and starspot groups within the orbital phase range that \citet{pil10,pil11,pil14,pil15} asserted was associated with enhanced X-ray flaring from a ``star-planet interaction.''  The results of our one-sample Kolmogorov-Smirnov test as applied to photometric maxima confirm that these apparent trends lack statistical significance.

These results are reinforced by those found in previous studies of photometry.  \citet{pon13} combined APT photometry with \emph{HST} infrared to UV transmission spectroscopy of the atmosphere of HD 189733b.  A byproduct of their analysis of exoplanet atmospheric characteristics was the ability to infer properties of stellar activity, including starspots and flares.  They noted that no monitoring campaign, whether ground- or satellite-based, had shown evidence for the transit of a significant bright region such as an accretion shock, although their analysis was limited to $\phi_{orbit}\sim$0.9-1.1.

\subsection{Lomb-Scargle Periodogram Analysis}
The Lomb-Scargle periodogram can also be leveraged for spectral analysis of the photometric maxima.  An advantage of this technique is the ability to search for periodic signals in unevenly sampled data while mitigating the potential determination of power in low-frequency signals that occur from large gaps in the data, such as occur between JD 2454329.80 and 2455130.00.  For this procedure, the data are sampled at $\sim$0.5 day intervals, in a binary fashion such that measured photometric maxima (Table 1, Column 3) are given values of 1, while non-maximal values in the temporal ranges bounded by Table 1, Columns 1 and 2 are given 0.  In this sampling scheme, the nearest 0.5 day interval to the maxima is removed and replaced with its measured photometric maximum time value.  Following the recommendation of \citet{pre07}, we choose an oversampling parameter of 4, and high frequency parameter $f_{hifac}=f_{hi}/f_{c}=2$, where $f_{hi}$ represents the highest frequency sampled, and $f_{c}$ is the Nyquist frequency.  This $f_{hifac}$ is selected so that we may search frequencies $f\leq$0.524 d$^{-1}$ ($P\gtrsim$1.910 d) that include the exoplanet orbital period, although we note that frequencies $f>$0.5 d$^{-1}$ may be aliased. 
	
The only significant frequency recovered by Lomb-Scargle is 0.16196478 day$^{-1}$ ($P=$6.1742 d) as shown in Figure 3.  This peak in the power spectrum has a false alarm probability of only 0.39\%, which denotes the probability that the null hypothesis, that the determined period is the result of independent Gaussian random values, is actually correct.  This value would appear to correspond to roughly half the rotation period of HD 189733A.  Interestingly, though, no peak in the power spectrum is found near the rotation period, perhaps denoting that at least two major sites of photometric maxima exist on the surface at any time, but that photospheric activity evolves somewhat over a single rotation.

A search of the power spectrum for a peak that corresponds to the exoplanetary orbital period ($P=$2.219 d, $f=$0.451 d$^{-1}$) finds no signal of statistical significance, to within double-point computational precision.  Similarly, although \citet{far10} suggested that the beat period of the rotational and planetary orbital periods may create a noticeable effect in the 2.5-2.7 d ($f=$0.370-0.400 d$^{-1}$) range, no statistically significant peak in the power spectrum was found at this frequency.

These analytical methods demonstrate that there is no persistent bright region consistent with a stream of continuously accreting gas that creates a shock and/or provides gaseous material that induces flaring, as \citet{pil15} assert.  Figure 2 also excludes a scenario in which intermittent gaseous accretion contributes significantly to the photometric properties of the HD 189733 A/b system.  Instead, the photometry demonstrate that any bright regions within $\phi_{orbit}\sim$0.52-0.65 are likely due to intrinsic stellar activity that is generally randomly distributed across the stellar photosphere.

\section{Discussion}

\subsection{Photometric Results within the Context of the T Tauri Paradigm}
Although the low cadence of \emph{most} of our photometry precludes the detection of AA Tau-like patterns, the higher-cadence \emph{MOST} photometry give no indication of such behavior.  Furthermore, we note that none of the photometric features, whether dark (ROME I) or light (this paper) are phased with the exoplanet revolution as proposed by \citet{pil15}; instead, they are all phased with the rotation period of the primary.  When phased with the rotation cycle, there are no stable bright regions that persist for more than a few rotations (Figure 1).  These results indicate that the bright regions denote intrinsic surface activity, such as plage, as opposed to accretion flows that reoccur at the same location over many years.  For comparison, \citet{don08} argued that the magnetic field geometry of the CTTS BP Tau consisted of an inclined 1.2 kG dipole and 1.6 kG octupole, accompanied by a dark spot formed near the accretion shock region.  This dark spot has a 2\% filling factor.  \citet{don11a,don11b} found similar cool, dark spots located at the footprints of polar accretion flows for the CTTSs V2129 Oph and TW Hya.  We note that our extensive analysis from ROME I precludes the existence of a long-lasting dark accretion spots on the surface of HD 189733A.  Thus, the morphology of the HD 189733 light curve clearly resembles a spotted naked photosphere, with no indication of a non-axisymmetric gaseous disk or trail of absorbing material linked to an accretion hotspot on the surface.

However, we acknowledge that non-steady accretion could produce short-lived and highly variable hotspots, as, for example, \citet{ale10} found in their \emph{Convection, Rotation and planetary Transits (CoRoT)} photometric survey of the young cluster NGC 2264.  The team found that $\sim$38\% of CTTSs exhibit irregular light curves that could be caused by variable accretion, variable hotspots, or random obscuration of the photosphere by non-axisymmetric circumstellar material.  They also determined that CTTS photometric variability ranged from 3\% to 137\%, far larger than the 1-2\% \emph{regular}, starspot-induced variability that \citet{pon13} measured at HD 189733A.  Thus, non-steady accretion may occur in the HD 189733 system at a much-reduced rate relative to CTTSs and contribute to the alleged ``on-off'' effect of enhanced magnetic activity.  However, this scenario is strikingly at odds with the assertion of a steady accretion stream \citep{pil15}, and leaves us with a mystery as to why no statistically significant congregation of photometric maxima are found in Figures 1-3.  We will investigate the observational evidence for variable accretion in more detail in Section 4.4.

\subsection{Classical T Tauri Emission at Radio Wavelengths}
Another lens through which to compare the HD 189733 and T Tauri systems is to examine their radio emissions, especially at 4-5 GHz (C-band).  Multiband radio observations of star forming regions have been conducted at 1.4, 4.8, 7.4, and 8.4 GHz \citep{rod83,one90,rod90,rod99,leo91,ruc92,tyc18}.  \citet{leo91} found that $\sim$33\% of sources within the $\rho$ Ophiuchi cloud had radio luminosities $\geq 2.5\times 10^{15}$ erg s$^{-1}$ Hz$^{-1}$ at 4.86 GHz and hypothesized that stellar magnetic activity was the principal source of this emission.  On the other hand, \citet{ang96} and \citet{rod99} hypothesized that bremsstrahlung radiation from ionized outflows cause radio jets that emit brightly at 4.9 and 8.4 GHz frequencies on account of the compact nature of the radio emission, their elongated spiral structures that aligned with young stellar object (YSO) outflow axes, and their relatively flat radio spectra.  \citet{and90} argued that 4.9 GHz radio emissions were thermal in origin due to their correlation with 1.3 mm emission.  More recently, the Gould's Belt Very Large Array Survey of the Ophiuchus, Serpens, Orion, Taurus-Auriga, and Perseus star-forming regions conducted at 4.5 and 7.5 GHz by the upgraded Karl G. Jansky Very Large Array (VLA) found that $\sim$60\% of YSOs have nonthermal, likely gyrosynchrotron, radio emission \citep{dzi13,kou14,dzi15,ort15,pec16}.  Alternatively, \citet{tyc18}, found that $\sim$55\% of Class 0 and Class I protostars emit bremsstrahlung radiation caused by J-type ionizing shocks during their VLA Nascent Disk and Multiplicity (VANDAM) survey at 4.7 and 7.4 GHz.

Several T Tauri systems have been the focus of more detailed study at radio wavelengths.  Multiband VLA observations of the CTTS HL Tau at 5, 8.3, and 23 GHz only appeared to reveal ionized, outflowing gas emitting via bremsstrahlung and found no evidence for nonthermal emission \citep{rod94,wil96}.  Very Long Baseline Interferometry (VLBI) observations of the CTTS system T Tau Bb at 8.3 GHz found $\sim$100\% right circularly polarized emissions that varied over timescales ranging from seconds to hours.  Their origin was attributed to a coherent emission process such as the electron cyclotron maser operating near magnetically confined accretion funnels \citep{smi03}.  Multiband VLA observations of the same system at 5, 8.1, and 15 GHz found $\sim$10\% left circularly polarized radio emission consistent with a gyrosynchrotron origin \citep{joh04,loi07}.

Using the prior results from T Tau Bb as a guide, we would expect an analogous source with its flaring radio luminosity ($L_{R}=1.38\times 10^{27}$ erg s$^{-1}$), located at the distance of HD 189733b, to have a 5 GHz radio flux density $F_{\nu}\sim$620 mJy.  On the other hand, cores in star-forming regions and CTTSs generally follow the G\"{u}del-Benz relationship which relates their X-ray luminosity to their incoherent radio luminosity (e.g., \citet{pec16}).  As we found in ROME I, the radio luminosity that corresponds to the X-ray flares measured by \emph{XMM-Newton} at HD 189733A is only $L_{R}=8.3\times 10^{22}$ erg s$^{-1}$.  Although the system had previously been observed at MHz radio frequencies, in addition to the 5-50$\times$ increased sensitivity of our exoplanet survey conducted at Arecibo Observatory, our survey was the first to observe HD 189733 at 4-5 GHz, and thus, the first to search for CTTS-like emission.  However, as reported in ROME I, we failed to detect variable radio emission similar to that observed from T Tau Bb during 2 hours of observations down to a 3$\sigma$, $\sim$1 s integration sensitivity of 1.158 mJy.  This corresponds to the ability to detect $\gtrsim$10\% circularly polarized flares as faint as $\nu~L_{\nu}\geq 2.183\times 10^{24}$ erg s$^{-1}$, or $\sim 10^{3}\times$ greater sensitivity than required to detect coherent emission from a T Tau Bb analog, but nearly an order of magnitude less sensitivity than required to detect gyrosynchrotron emission in accordance with the G\"{u}del-Benz relationship.  Similarly, although bremsstrahlung radiation been detected from T Tau Bb at 149 MHz with the Low Frequency Array (LOFAR) \citep{cou17}, 148, 244, 327, and 614 MHz observations of HD 189733 conducted earlier did not result in any detections, despite the sensitivity to do so (see ROME I and references therein).  Thus, HD 189733 clearly differs from CTTS systems at radio wavelengths in that it is underluminous and lacks certain types of emission processes.

\subsection{Accretion Constraints Derived from Lyman-$\alpha$ Measurements of Exospheric Evaporation}
If HD 189733A is accreting matter from the exoplanet that may fuel enhanced magnetic activity, one of the critical parameters to determine is the mass accretion rate.  \citet{lec10} first found evidence for an evaporating exosphere for HD 189733b during \emph{HST} Advanced Camera for Surveys (ACS) observations of three transits in Lyman-$\alpha$ (1215.6\AA) in 2007-2008.  They measured a transit depth of 5.05$\pm$0.75\%, whereas the depth associated with a planetary disk alone should be $\sim$2.4\%.  This Lyman-$\alpha$ absorption depth and its doppler-broadened properties can be explained by either hydrogen gas overfilling the Roche lobe, or that the gas exceeds the escape velocity of the planet, both of which indicated the existence of an evaporating exosphere for HD 189733b.  They then simulated the exoplanet transit depth and its temporal evolution as a function of the exoplanet escape rate and the EUV flux.  They found that the data were best fit by escape rates of $\dot{M}=10^{9}$ to 10$^{11}$ g s$^{-1}$ accompanied by EUV fluxes of 10-40$\times$ solar values ($F_{\sun}$), with an optimal fit of $\dot{M}=10^{11}$ g s$^{-1}$ and $F_{EUV}=20 F_{\sun}$.

Follow-up \emph{HST} Space Telescope Imaging Spectrograph (STIS) EUV observations found highly variable UV absorption, with no exospheric absorption detected in 2010, but a Lyman-$\alpha$ absorption depth of 14.4$\pm$3.6\% detected in 2011 \citep{lec12}.  Models simultaneously fit to their \emph{HST}/STIS and \emph{XMM-Newton} 0.3-3keV band data indicated an atomic hydrogen escape rate of $\dot{M}=4.4\times10^{11}$ g s$^{-1}$ and $F_{EUV}=5 F_{\sun}$.  The authors noted that this value represents a lower limit, as it does not include the escape of ionized hydrogen.  These observations demonstrate the great temporal variability of exospheric evaporation from HD 189733b that may be intimately linked to stellar activity.  From these three observation epochs, \citet{pil15} estimated a value of $\dot{M}=5\times 10^{18}$ g yr$^{-1}$ (or $\dot{M}=1.58\times 10^{11}$ g s$^{-1}$ averaged over an entire year) for use in their MHD simulations of accretion from the exoplanet, a value that we adopt here as well.  

Given this accretion rate, \citet{pil11,pil14,pil15} linked their results with accretion from T Tauri stars.  However, an average accretion rate of $\dot{M}=10^{9}$ to 10$^{11}$ g s$^{-1}$ is much smaller than those found for CTTSs, which have a typical accretion rate $\dot{M}\sim~10^{18}$ g s$^{-1}$, but range from $\dot{M}\sim~10^{13}$ to $10^{21}$ g s$^{-1}$ depending on stellar mass, age, and intrinsic variability \citep{har16}.  Rather, after a careful examination of the literature on solar and stellar phenomenon, we determined that the closest analogy for the estimated accretion rate at HD 189733A is outgassing from sungrazing comets.  For example, \citet{mcc13} leveraged \emph{Solar Dynamics Observatory (SDO)} Atmospheric Imaging Assembly (AIA) UV and \emph{Hinode} X-Ray Telescope (XRT) data to derive an outgassing rate of $\sim9.5\times 10^{9}$ g s$^{-1}$ during outburst from Comet Lovejoy (C/2011 W3).  In total, the comet is estimated to have lost $\sim10^{13}$ g during the egress phase of its perihelion passage alone. Strikingly, neither the careful photometric analysis of \citet{mcc13}, nor the spectroscopic analysis of \citet{ray18} reported evidence for the initiation of solar flares during the comet's passage within 0.2R$_{\odot}$ of the solar photosphere.  To our knowledge, there are no reports in the literature of the passage of sungrazing comets triggering solar flares or coronal mass ejections (CMEs).  Thus, we can safely assume that this meager accretion rate is far too small to trigger magnetic reconnection within a stellar corona, as \citet{pil15} suggested occurs in the vicinity of the accretion column at HD 189733A.

\subsection{Optical and Ultraviolet Spectral Line Diagnostics of Accretion}
The analysis of the velocity characteristics of optical and UV spectral lines provide strong constraints on the gas dynamics within the HD 189733 system and aid in its comparison to CTTSs.  \citet{pil15} supplemented their X-ray observations of the HD 189733 system with \emph{HST}/COS observations in 2013.  They used the G130M grating to monitor the FUV range 1150-1450$\AA$ during orbital phases $\phi_{orbit}$=0.500-0.626.  They observed two flares at $\phi_{orbit}$=0.525 and $\phi_{orbit}$=0.588 in the lines \ion{C}{2} (1334.5-1335.7\AA), \ion{N}{5} (1238.8-1242.8\AA), \ion{Si}{2} (1264.7\AA), \ion{Si}{3} (1201.0-1206.6\AA), and \ion{Si}{4} (1393.8,1402.8\AA).  During the first flare, they computed that the line centroids for various species of \ion{C}{2}, \ion{Si}{2}, \ion{Si}{3}, and \ion{Si}{4} were redshifted by $\Delta v\lesssim +$18$\pm$5 km s$^{-1}$ with full widths at half maxima (FWHMs) of $\sim$40km s$^{-1}$.  The line centroids for the same species were blueshifted in the second, weaker flare by $\Delta v\gtrsim -$23$\pm$5 km s$^{-1}$ again with FWHMs$\sim$40km s$^{-1}$.  Through analysis of the differential emission measure, the team determined that the second event was more impulsive and hotter, but both events had plasma temperatures in the range $T=(1-80)\times$ 10$^{4}$ K, which is consistent with the transition region.  \citet{pil15} attributed these characteristics to structures in a stream of gas steadily accreted from the evaporating exosphere of HD 189733b.  They suggested that these structures and a surface accretion spot rotate at the orbital velocity of the exoplanet (17.3 km s$^{-1}$), as opposed to the rotational speed of the star (3.2 km s$^{-1}$), despite this scenario being inconsistent with the observed behavior of gas within the vicinity of accreting CTTSs.

The \emph{HST}/COS spectra offer several curious characteristics.  Although several FUV lines were analyzed, only the \ion{Si}{4} lines appear to be properly calibrated with quiescent measurements centered around zero and velocity centroid shifts of similar magnitude to the orbital motion of the planet.  The other lines have non-zero centroids; in particular, the \ion{C}{2} and \ion{Si}{3} line centroids exhibit a systematic offset toward redder wavelengths, likely as a result of \emph{HST}/COS pointing, wavelength calibration, and geometric correction errors which have been estimated to introduce velocity uncertainties $\sim$7 km s$^{-1}$ as described in detail in \citet{ard13}.  Removal of these biases obviously reduces the magnitude of the redshift of the emitting gas, while increasing the magnitude of the blueshifted components, thereby making the results inconsistent with a scenario of accreted material traveling across the stellar surface at the magnitude of the exoplanet orbital velocity that results in symmetric blue and redshifts.  The velocity centroids for the \ion{N}{5} line at 1242.8$\AA$ are all blueshifted, as are 5/6 of the subexposures for the 1238.8$\AA$ line, which is inconsistent with redshifted accretion activity, but is consistent with a stellar wind.  We also note that the FWHMs of the spectral lines are less than half those generally measured in accreting CTTS systems (FWHM$\gtrsim$100 km s$^{-1}$), again suggesting the absence of an accretion stream with significant velocity gradients (e.g., \citet{ard13}).  The impulsive nature of the spectral line flux increases are rather difficult to square with a shock or ``knee'' feature in an accretion stream that is unobscured for half of the stellar rotation period.

The most important FUV spectral lines for assessing the activity of T Tauri stars have been \ion{C}{4} (1548.2, 1550.8 \AA) and \ion{He}{2} (1670.5\AA) \citep{ard13}.  Previous work has found a correlation between \ion{C}{4} luminosity and accretion rate in CTTSs, while this species is uncorrelated with stellar activity.  \ion{He}{2} is a tracer of post-shock accretion.  Although \citet{ard13} relied on both \emph{HST}/COS and \emph{HST}/STIS to conduct spectroscopy of CTTSs, WTTSs, and transition disk objects in the FUV range 1150-1790\AA, it is unfortunate that \citet{pil15} did not observe these same lines with the same instrumentation.

Some of the key diagnostics of magnetospheric accretion among CTTSs at optical wavelengths are the \ion{He}{1} lines at 5875.6$\AA$ and 6678\AA.  These emission lines are composed of broad and narrow core components; the broad core is likely derived from either a stellar wind or a large-scale, magnetically controlled accretion flow that spans from the inner truncation radius of the circumstellar disk to the stellar photosphere, while the narrow core is thought to be generated from an accretion shock near the photosphere.  High-resolution spectropolarimetry of the narrow cores of these spectral lines revealed magnetic field strengths of $B_{Z}\sim$-2 kG at TW Hya and $B_{Z}\sim$6 kG at GQ Lup \citep{joh13}.  Polarimetry of the \ion{He}{1} 5876$\AA$ line at TW Hya also determined that the accretion flows are modulated by the rotation of the pre-main-sequence star \citep{yan07,don11b}.  These results demonstrate that CTTS accretion structures rotate with the star since the plasma is ionized and therefore is controlled by the strong magnetic fields in the stellar magnetosphere.  Although the magnetic field of HD 189733A is generally weaker than in CTTSs, photospheric accretion spots and coronal structures should still engage in nearly solid body rotation.  This behavior differs markedly from the scenario that \citet{pil15} proposed, where various accretion structures move in opposing directions, but with maximum velocity equivalent to the orbital speed of the exoplanet.  A consequence of their scenario might be that any \ion{He}{1} 5876$\AA$ line emission is modulated at the timescale of the exoplanet orbital period.  However, \citet{boi09} found that this spectral line was only rotationally modulated, without any abnormal excess emission at $\phi_{orbit}\sim$0.52-0.65, as might be anticipated (see their Figures 5 and 6).  We note that \ion{He}{1} is also an indicator of chromospheric activity such as that found near active regions in main sequence stars and need not be associated with accretion at all.  A small contribution to the \ion{He}{1} line emission from an accretion process would likely be dwarfed by intrinsic chromospheric activity.

As stated in the introduction, properties of the H$\alpha$ emission of pre-main-sequence stars have been important to understanding and classifying their accretion and intrinsic activity.  \citet{cau15} conducted transmission spectroscopy during two transits of HD 189733b using HiRES on Keck I, and analyzed the temporal evolution of the equivalent widths of H$\alpha$, H$\beta$, and H$\gamma$.  They found evidence for pre-transit absorption in all three spectral lines $\sim$120 min and 70-55 min before transit, but not post-transit.  They argued that since the absorption centroids were shifted no more than $\pm$5 km s$^{-1}$, as opposed to significantly redshifted absorption as would be expected of an accretion stream, that their data were consistent with a magnetospheric bow shock at a distance of 7.2 planetary radii ($R_{P}$).  Further work confirmed pre-transit absorption in H$\alpha$ and H$\beta$ $\sim$220 to 150 min before transit, corresponding to a distance of $\sim$17 R$_{P}$ ahead of the exoplanet \citep{cau16}.  Since the newer observations disagreed with the previous bow shock model, the team explored the possibility that the pre-transit absorption signatures might be the result of a clumpy accretion stream derived from the evaporating exosphere.  However, they decided against such a scenario because it would result in +70-100 km s$^{-1}$ redshifts, far larger than the +5-10 km s$^{-1}$ line centroids observed.  In addition, the presence of an ionized stellar wind and significant photoionization would make it difficult to maintain a significant neutral hydrogen population so far from its source.  We note that the team inferred clump masses of $\sim$10$^{7}$ g from the optical spectra, which are far smaller than the measured Lyman-$\alpha$ evaporation rates.  However, they acknowledged that the pre-transit absorption signatures could be explained by even small levels of stellar activity.  Continued investigations of the circumstellar environment indicated that evaporated material could orbit $\sim$5 R$_{P}$ ahead of the planet in their 2013 June data set, but a similar distance behind the planet in their 2006 August data set \citep{cau17}.  The team argued that only a fine-tuned stellar activity model could account for the H$\alpha$ absorption variability observed, thus leading them to favor an interpretation related to the planetary atmosphere.  Once again, the existence of accretion streams could be excluded from their compiled data because they measured small, $\pm$7 km s$^{-1}$ H$\alpha$ velocity centroids.  

Within the CTTS paradigm, these results differ significantly from those found in accreting systems.  For instance, \citet{don11b} measured H$\alpha$ and H$\beta$ strongly in emission at TW Hya, with equivalent widths $\sim$200$\AA$ and $\sim$30$\AA$, respectively.  By comparison, in Table 2 summarizing their observations, \citet{cau17} measured $W_{eq}$(H$\alpha$)$\sim$10$^{-2}\AA$ across all observations, revealing that any accretion flow is extremely tightly confined within velocity space, or has yet to be detected.  Thus, although these pre-transit observations appear to support the hypothesis of a stream of matter accreting as hypothesized by \citet{pil10,pil15}, the H$\alpha$ velocity centroid and equivalent width measurements pose significant challenges.

On the other hand, these spectroscopic properties are entirely consistent with stellar flares.  While \citet{pil15} looked to the analysis of solar flare fallback material in \citet{rea14} to gain a greater understanding of how an accretion stream might appear at HD 189733A, we note that their case study offers a simpler explanation that better fits the data: typical stellar flares.  The impulsive nature of the flares is consistent with solar flare morphologies, and as \citet{ben13} noted, changes in solar chromospheric activity result in stronger variation in the \ion{Si}{4} and \ion{Si}{3} lines than \ion{O}{1} and \ion{C}{2} lines, as observed at HD 189733A.  In their review of the observational characteristics of solar flares, \citet{fle11} found that upflow and downflow velocities of flare material within the chromosphere and transition region ($T\sim 10^{4}-10^{5}$K) vary from -20 to +50 km s$^{-1}$, neglecting projection effects.  This velocity range matches those measured by \citet{pil15} and \citet{cau17} at FUV and optical wavelengths, respectively.

\subsection{Is HD 189733A A T Tauri Analogue at X-Ray Wavelengths?}
\subsubsection{Flare Temperature and Density Characteristics}
Next, we examine the data, methods, and results of the analysis of the X-ray flares observed from 2007 to 2012 by \citet{pil10,pil11,pil14} so that we may characterize the plasma densities and temperatures related to the hypothesized accretion activity and make inferences about the physical processes that caused the flaring.  Our reanalysis will leverage updated plasma ionization tables, investigate the contribution to X-ray emission from flaring and quiescent states, and consider the X-ray activity of HD 189733A within the context of other magnetically active stars.

\citet{pil10} reported on two epochs of observations of HD 189733 with \emph{XMM-Newton}, one during planetary transit in 2007, and one during secondary transit in 2009.  They stated that no flares were observed during the first 54 ks session, while a $\sim$9 ks complex flare was observed at  $\phi_{orbit}$=0.54 during the latter 35 ks session.  \citet{pil11} witnessed a $\sim$11 ks complex flare at $\phi_{orbit}$=0.52 during a 39.1 ks observation in 2011.  The team also observed a $\sim$10 ks complex flare at $\phi_{orbit}$=0.64 during a 61.5 ks session in 2012 \citep{pil14}.  For all data sets, they collected light curves and spectra with the European Photon Imaging Cameras (EPIC) and spectra with the Reflection Grating Spectrometer (RGS).  Using this information, in each paper they investigated plasma temperatures, emission measures, plasma densities, gas kinematics, and X-ray fluxes, where the later papers revised the analysis from the previous papers.  Therefore, we will direct our discussion to the results of the latest paper in the series, except where there was some analysis that is pertinent to our discussion but has not been updated.

\citet{pil14} fitted two or three temperature Astrophysical Plasma Emission Code (APEC) models to spectra from 2007, 2009, 2011, and 2012 epochs.  The spectra in each epoch (except for 2007 when the team reported no flaring activity) were divided into pre-flare, flare, and post-flare phases, with flare and post-flare phases allowed to consist of three temperature models that included an additional component to describe the potentially exoplanet-induced flares.  When three-component models were employed, the first two plasma components were fixed at the values determined from the pre-flare state in each epoch.  From these, they found that HD 189733A's corona is well described by a cool plasma component at $kT_{1}=$0.18-0.24 keV and a warm component at $kT_{2}=$0.47-0.73 keV.  During the flares in 2009, 2011, and 2012, their models indicated the presence of a third, hotter plasma temperature at $kT_{3}=$0.82-0.99 keV ($T=9.5\times$ 10$^{6}$ to $T=1.1\times$ 10$^{7}$K).  In all cases, the hotter component disappeared by the post-flare phase (see their Table 1).

Coronal densities were computed for HD 189733A in both \citet{pil11} and \citet{pil14}.  In the former, the $r$, $i$, and $f$ line ratios were measured for \ion{O}{7} and \ion{Ne}{9} in both quiescent and flaring states and used to compute $R=f/i$.  These spectral features were each detected in only one of the two RGS detectors.  While the statistics of the lines during quiescence were generally robust, those measured during flaring episodes tended to have average fluxes within 2-3$\sigma$ of zero. The relatively large uncertainties and difficulty in measuring the intercombination lines required optimistic evaluation to yield physically meaningful values.  During quiescence, they determined electron densities of $3.2\times 10^{10}$ cm$^{-3}<n_{e}<1.3\times 10^{11}$ cm$^{-3}$ and $n_{e}<1\times 10^{11}$ cm$^{-3}$ via the \ion{O}{7} and \ion{Ne}{9} $R$ ratios, respectively.  During flaring, they inferred $n_{e}<1.6\times 10^{11}$ cm$^{-3}$ from \ion{O}{7}, while the \ion{Ne}{9} lines offered no constraints.  They surmised from these values that the coronal density decreases during flares as compared to the quiescent state.  We note, however, that the derived quiescent and flaring coronal densities are equivalent within observational uncertainty.  They concluded that the corona of HD 189733 is denser, 10 times more luminous, and more active than main sequence stars without ``hot Jupiter'' companions, which could be due to star-planet interactions.

\citet{pil14} further investigated the coronal density by measuring \ion{O}{7} $R$ ratios aggregated from all of their observations spanning $\sim$190 ks to improve the signal-to-noise ratio of their analysis.  These ratios included photons that arrived during both flaring and quiescent phases from which they calculated $n_{e}=(3-10)\times 10^{10}$ cm$^{-3}$ (68\% confidence), or $n_{e}=(1.6-13)\times$ 10$^{10}$ cm$^{-3}$ (90\% confidence). They remarked that HD 189733A hosts a corona that is slightly denser than those found on less active stars.

Using the He-like triplet line ratios, we can independently constrain the coronal density of HD 189733A by leveraging updated calculations of hybrid photoionized and collisionally ionized plasmas \citep{por01,gud09}.  This analysis will consist of two parts: using the $R$ and $G=(f+i)/r$ line ratios to make inferences about the coronal density and temperature, and an evaluation of the repercussions of combining the line ratios of all data sets. The $R$ line ratio can be approximated by 
\begin{equation} R={f \over i} = {R_{0} \over 1+n_{e}/N_{C}},\end{equation}
where $R_{0}$ is the low density limiting flux ratio and $N_{C}$ is the critical density at which $R=R_{0}/2$.  Using the $R_{0}$ and $N_{C}$ values presented in \citet{gud09}, Table 1 we can estimate electron densities for the fluxes found in \citet{pil11,pil14}.  We derive $n_{e}\sim8.9\times 10^{10}$ cm$^{-3}$ and $n_{e}\sim1.1\times 10^{10}$ cm$^{-3}$ from \ion{O}{7} in quiescence and flaring states, respectively, and $n_{e}\sim5.7\times 10^{10}$ cm$^{-3}$ for the entire ensemble of \ion{O}{7} flux ratios.  The \ion{Ne}{9} quiescent flux measurements yield unphysical results due to the poor quality of the measurements, but during flaring we obtain $n_{e}\sim1.2\times 10^{12}$ cm$^{-3}$.  We note that the flaring density derived from \ion{Ne}{9} is $\sim$100$\times$ greater than that calculated from \ion{O}{7}.  Our approximate density values can be compared to those obtained from the the ``rline'' table constructed by \citet{por01}, assuming a radiation field factor of $W=\onehalf$ as occurs for coronal structures near the photosphere.  In this instance, the \ion{O}{7} $R$ ratios imply $n_{e}=(2-10)\times 10^{10}$ cm$^{-3}$ in quiescence and $n_{e}=(1-10)\times 10^{9}$ cm$^{-3}$ in flaring, while the \ion{Ne}{9} $R$ ratios imply $n_{e}=(5-10)\times 10^{11}$ cm$^{-3}$ during flares.  Interestingly, using the ``gline'' table of \citet{por01} yields surprisingly cool plasmas of $T<2\times$ 10$^{6}$ K in both quiescent and flaring states, nearly an order of magnitude less than that derived from the hottest component in the APEC models.  While $T\sim(1-6.3)\times$ 10$^{6}$ K plasma is found in solar quiet regions and the hot corona, flares generally have temperatures $T\sim(1-4)\times$ 10$^{7}$ K \citep{asc05}.  If we exclude the \ion{Ne}{9} density analysis due to its poorer statistics, we find that the corona is less dense while flaring and denser in quiescence as \citet{pil11} found, although the plasma temperatures that we compute based on the line ratios are significantly smaller than those indicated by the APEC models.  However, given the large observational uncertainties in the \ion{O}{7} and \ion{Ne}{9} line ratios, it is difficult to form firm conclusions about the plasma densities and temperatures.  Table 2 summarizes these temperature and density results.

A potential concern among the analysis methods used to improve the X-ray spectra signal-to-noise ratio is that \citet{pil14} combined \emph{all} of the \emph{XMM-Newton} RGS spectra available, which consists of 54, 35, 39.1, and 61.5 ks of observations from the 2007, 2009, 2011, and 2012 campaigns, respectively \citep{pil10,pil11,pil14}.  These data sets contain X-ray photons from HD 189733A in \emph{both} flaring and quiescent states, such that the corresponding flare durations for those observations are $\sim$4, 9, 11, 2, and 10 ks as summarized in ROME I, Table 1.  These flare durations would suggest that a more active corona is observed for 36 ks out of a total \emph{XMM-Newton} observation duration of 189.6 ks, or $\sim$19\% of the time.  Therefore, the computed coronal density actually represents a mixture of flaring and quiescent states, with their differing densities.

An examination of the solar coronal density proves instructive in evaluating the claim of coronal overdensity at HD 189733A.  We can estimate the average coronal electron density, $n_{e}$, for HD 189733A using the equation:
\begin{equation} n_{e}={\left(\tau_{flare}\over \tau_{total}\right)}{\left[(f_{flare flux})n_{e,flare}+ (1-f_{flare flux})n_{e,quiescent}\right]}+{\left[1-{\left(\tau_{flare}\over \tau_{total}\right)}\right]}n_{e,quiescent},\end{equation}
where $\tau_{flare}$ denotes the total time during the four observations when HD 189733A was in the high flaring state, $\tau_{total}$ denotes the total length of the \emph{XMM-Newton} observations, $f_{flare flux}$ denotes the fraction of the star's X-ray flux that is due to the eruption of large flares, $n_{e, flaring}$ denotes the solar coronal electron density during flares, and $n_{e, quiescence}$ denotes the quiescent solar coronal electron density.  In 2007, the flare at $\sim$5 ks increases the X-ray flux from $\sim$60 to 90 cts s$^{-1}$.  In 2009 and 2011, the count rates associated with flaring activity increased from $\sim$110 to 200 cts s$^{-1}$ and $\sim$110 to 220 cts s$^{-1}$, respectively.  Therefore, the maximum contribution to the observed X-ray flux from the large flares is $\sim$50\% ($f_{flare flux}\sim0.5$).  If we assume solar values of $n_{e, quiescence}\sim$10$^{9}$ cm$^{-3}$ (above quiet regions) and $n_{e, flaring}\sim$10$^{11}$ cm$^{-3}$ (within flare loops), we compute the coronal density of HD 189733A to be $n_{e}\approx$10$^{10}$ cm$^{-3}$ from solar values alone \citep{asc05}.  Thus, by accounting for the temporal weighted average of quiescent and flaring states, our estimate is within 2$\sigma$ of the value \citet{pil14} computed, without requiring an overdense corona at HD 189733A.

To conclude, we have used updated calculations of the \ion{O}{7} and \ion{Ne}{9} triplet line ratios to attempt to better constrain the coronal density of HD 189733A.  However, the effort is largely confounded by measurement uncertainty, despite various tools and methods to improve the statistics.  Although $\sim$190 ks of spectral data have been collected, because these data reflect a fusion of flaring and quiescent states, they do not shed light as to whether the corona of HD 189733A is surprisingly dense.  Our calculations indicate that the observed coronal properties are, in fact, consistent with a corona of nearly solar density to within measurement uncertainty.  However, even if HD 189733A has a denser-than-solar corona, higher coronal densities of $n_{e}\sim$10$^{10}$-10$^{11}$ cm$^{-3}$ are typical for magnetically active stars, making an explanation that relies on accretion unnecessary \citep{gud09}.  These calculations merely support the analysis in ROME I that HD 189733A is an active star.

\subsubsection{Potential Accretion Shock Properties}
Protostars emit X-rays by a variety of mechanisms.  They are strong sources of coronal magnetic activity, with plasma temperatures $T\sim 10^{7}$ K.  In accreting CTTS systems, plasma from beyond the corotation radius, or inner truncation radius of the circumstellar disk, is accelerated at supersonic velocities toward the stellar photosphere and shocked, where it is heated to $T\sim 10^{6}$ K.  An especially deep ($\sim$489 ks) observation of the CTTS TW Hya with the \emph{Chandra X-ray Observatory} High Energy Transmission Grating (HETG) provides an excellent case study of these phenomena.  Based on the $G$-ratios of \ion{O}{7}, \ion{Ne}{9}, and \ion{Mg}{11}, \citet{bri10} measured $T_{e}=(2.1-2.5)\times 10^{6}$  (0.22 keV) for the accretion shock region.  Using the $R$-ratios for these He-like ions, they found $n_{e}=(5.7-58)\times 10^{11}$ cm$^{-3}$.  They attributed \ion{Ne}{9} and \ion{Mg}{11} to the shock region and \ion{O}{7} to the post-shock region (although \ion{O}{7} is also formed by coronal processes).

These results yield insights into the X-ray activity observed at HD 189733A.  First, the plasma temperatures measured in the soft X-ray emitting shock regions of TW Hya are a factor of $\sim$4 smaller than those computed for the largest flares at HD 189733A by the APEC models \citep{pil14}, although they are consistent with the temperatures derived from \ion{O}{7} $G$-ratios in the previous section.  Second, the plasma density measured in the TW Hya accretion shock is a factor of 40-200$\times$ greater than that derived above by \citet{pil14}.  Third, \citet{pil11} indicated that they observed a softening of the X-ray spectra during planetary (secondary) eclipse in their 2009 data, but not in their 2011 data.  Indeed, the hardening of the PN light curve that they depict in their Figure 1 is entirely consistent with coronal flaring, yet inconsistent with the softening of X-ray emission expected in the presence of an accretion stream and shock.

We can also use the accretion rate found in Section 4.3 to determine the luminosity and temperature properties of the shocked gas in the purported ``knee'' feature.  As described above, CTTS accretion shocks usually emit in soft X-rays.  However, the accretion rate of HD 189733A is considerably different than that of CTTSs, potentially leading to shock luminosities and temperatures that differ significantly from the T Tauri star paradigm.  The maximum accretion luminosity can be calculated assuming the complete transformation of gravitational potential energy into the kinetic energy of accretion \citep{har16}:
\begin{equation} L_{S}={1 \over 2} \dot{M} v_{ff}^{2}={G M_{\ast} \dot{M} \over R_{\ast}} \left(1-{R_{\ast} \over R_{M}} \right),\end{equation}
where $\dot{M}$ is the accretion rate, $v_{ff}$ is the freefall velocity, $M_{\ast}$ is the primary star mass, $R_{\ast}$ is the primary star radius, and $R_{M}$ is the magnetospheric radius in the T Tauri model, but here we replace it with the semimajor axis of the exoplanet orbit.  This yields an accretion luminosity of $L_{S}=2.88\times 10^{26}$ erg s$^{-1}$ that is a factor of $\sim 10^{2}$ below the detection threshold of the existing \emph{XMM-Newton} observations and quite undetectable by our photometric study.

Using the strong shock approximation, we can also derive the shock temperature \citep{har16}:
\begin{equation} T_{S}={3 \over 16} \left( {\mu m_{H} \over k}\right) v_{ff}^{2},\end{equation}
where $\mu=0.5$ for an accretion flow consisting entirely of ionized hydrogen, and $m_{H}$ is the standard atomic mass of hydrogen.  This yields a plasma temperature $T_{S}=4.14\times 10^{6}$ K, which peaks at $\lambda_{max}$=7\AA.  These calculations indicate that shocked plasma in the HD 189733 system will be somewhat hotter than that observed in a CTTS system like TW Hya, but the X-ray spectrum of the shock should still be observed to soften.  Moreover, the computed maximum accretion luminosity poses particular difficulties for the accretion model of \citet{pil14,pil15}, since it implies an undetectably weak accretion flow.

\subsection{Flare Length Models}
An important line of evidence to support the hypothesis of a recurring star-planet interaction between HD 189733A and its exoplanetary companion came from analysis of the temporal variability of the 2012 X-ray flare light curve, and the subsequent apparent discovery of extremely large flares on the primary \citep{pil14}.  In this section, we will investigate whether the flares observed at HD 189733A are anomalously large and therefore consistent with magnetic reconnection occurring in an accretion stream.  First, we will leverage several methods to estimate the flare length, including by refining and updating the analysis used in \citet{pil14}.  Second, we will compare the flare lengths that we derive to those found on other stars without ``hot Jupiter'' companions.

Using the model of \citet{zai89} that X-ray magnetoacoustic oscillations could be triggered by a centrifugal force acting on evaporating plasma confined to a magnetic loop, \citet{mit05} sought to derive the flare loop length and magnetic field strength.  They studied the \emph{XMM-Newton} light curve of AT Mic (GJ 799A/B), a dM4.5e+dM4.5e binary system and derived loop lengths, $L$, for slow (acoustic), fast kink, and fast sausage magnetoacoustic wave modes.  For standing waves,
\begin{equation} L = jc\tau/2,\end{equation}
where $j$ is the mode number, assumed to be 1 for the fundamental mode, $c$ is the wave speed for the appropriate mode, and $\tau$ is the period of the oscillation.  The wave speeds of the various modes are related to the sound speed, $c_{s}$, by a factor of order unity.  \citet{pil14} had characterized the flare plasma with electron density, $n_{e}=(1.6-13)\times 10^{10}$cm$^{-3}$, temperature $T=1.1\times 10^{7}$ K, and oscillation periods of $\tau=4\pm1$ ks and $\tau\sim 9$ks, with the latter period being less significant.  When these values, including the longer oscillation period, are placed into the equation for the sound speed, assuming an adiabatic gas ($\gamma=5/3$):
\begin{equation} c_{s}=\sqrt{{2\gamma~k_{B}~T \over m_{p}}},\end{equation}
we recover a sound speed $c_{s}=5.5\times 10^{7}$ cm s$^{-1}$ as \citet{pil14} did, although we note that their equation lacks the square root.  \citet{pil14} reported that, \emph{``According to this result, the loop length is $L\simeq 4\times 10^{11}$ cm or approximately 4$R_{\ast}$, assuming $R_{\ast}=0.8R_{\sun}$.  This gives a height for the loop of $\sim$0.007 AU...this length constitutes a significant fraction of the distance between the two bodies ($\sim$25\%).''}  However, it is here that our derivations diverge as we are unable to reproduce the physical length scales that they describe.  If we include the factor of $\onehalf$ attributable to standing waves, we determine a loop height $L=2.5\times 10^{11}$ cm for  the case $\tau=$9000 s.  When converted into stellar radii ($R_{\ast}=0.76R_{\sun}=5.3\times 10^{10}$ cm), this yields $L=4.7R_{\ast}$, or 0.017 AU.  Given that the exoplanet has a semimajor axis $a=4.6\times 10^{11}$ cm, the flare would stretch $\gtrsim$50\% of the distance between the two bodies.  We note, however, that from \citet{pil14}, Figure 7 an oscillation period of $\tau\sim 9$ ks is not supported by the wavelet analysis, although $\tau\sim 6.5$ ks is supported with less significance than the shorter oscillation.  Using the most statistically sound period, $\tau=$4000 s, we compute a flare length $L=1.1\times 10^{11}$ cm, which is equivalent to $L=2.1R_{\ast}$, or 0.007 AU and still yields a relatively large flare.  In addition, when \citet{pil14} used the formula found in \citet{mit05} for the period of a slow-mode, standing wave, they found $L\sim1\times 10^{11}$ cm for $\tau=$4000 s, in rough agreement with their earlier calculations, despite the differences in periods involved and conversion errors.

Other methods may be used to estimate flare lengths that do not assume that a standing magnetoacoustic wave is present.  In the case of pressure balance between the evaporated plasma gas and the magnetic pressure, the flare loop length can be computed via
\begin{equation} L=10^{9}\left({EM\over 10^{48}cm^{-3}}\right)^{3/5} \left({n_{e}\over 10^{9}cm^{-3}}\right)^{-2/5} \left({T\over 10^{7}K}\right)^{-8/5} [cm],\end{equation}
\citep{shi02,mit05}.  Using $n_{e}=1.6\times 10^{10}$ cm$^{-3}$ and emission measure $EM=1.6\times 10^{51}$ cm$^{-3}$, we derive $L=6.0\times 10^{10}$ cm.  If instead we use the upper bound for the plasma density of $n_{e}=1.3\times 10^{11}$cm$^{-3}$, we find $L=2.6\times 10^{10}$ cm.  Note that we assume that the emission measures recorded in \citet{pil14}, Table 1 actually have units of 10$^{51}$ cm$^{-3}$, as they do in \citet{pil10}.

Alternatively, the flare length may be estimated from the rise and decay times of the complex flare temporal profile under the assumption of optically thin, ``coronal'' radiative cooling \citep{haw95,mit05}:
\begin{equation} L={1500\over \left(1-x_{d}^{1.58}\right)^{4/7}}~\tau_{r}^{3/7}~\tau_{d}^{4/7}~T^{1/2} [cm].\end{equation}
From \citet{pil14}, Figure 7 we derive a rise time, $\tau_{r}\sim$2500 s and a decay time, $\tau_{d}\sim$6000 s.  Note that $x_{d}^{2}=c_{d}/c_{max}$, where $c_{d}$ is the count rate at the end of the flare, for which we use the baseline rate before and after the flare, and $c_{max}$ is the peak count rate.  Assuming radiative cooling, we derive $L=3.0\times 10^{10}$ cm.  Thus, these three methods of calculating the flare length differ by approximately an order of magnitude, with methods that assume standing wave oscillations delivering higher values, while those based on pressure balance and cooling times yield smaller and more reasonable values (see Table 3).  However, our analysis restricts the range of flare lengths to a more internally consistent range of $L\sim3\times10^{10}$ to $1\times10^{11}$ cm.  We note that flare pulsation times tell us nothing about the viewing geometry of the activity, leaving us to conservatively assume an entirely radial flare for easier comparison with the semimajor axis of the exoplanet.  This, of course, need not be the case, highlighting that our estimates establish the upper limit on how far flares may stretch above the photosphere, while in reality their radial extent may be significantly less.

Although we have examined the flare lengths in terms of the stellar and exoplanetary parameters of the HD 189733 system, it would also be helpful to compare these flare lengths and timescales with those measured on other stars.  The best studied cases are of course solar flares, with typical length scales of $L\sim0.3 R_{\sun}=2.1\times 10^{10}$ cm \citep{def07}.  Quasi-periodic pulsations (QPPs) are found among solar flares with typical periods that span sub-second to $\sim$5 min timescales \citep{doo16}.  Perhaps the longest QPP recorded on the Sun, during the 2003 February 5-6 sequences of flares, was observed at multiple wavelengths stretching from radio (1-2, 17 GHz), to H$\alpha$, and X-rays (1-8 \AA, 0.5-4 \AA, and 3-25 keV).  \citet{fou05,fou10} determined that the $\sim$18 min QPPs at 1-2 GHz ($\sim$10 min QPPs at all other wavelengths) of this flare could be attributed to fast kink mode oscillations of a loop estimated to be $L=5.00\times 10^{10}$ cm to $1.410\times 10^{11}$ cm, without invoking any star-planet interaction.  Similarly, \citet{mae15}, in their study of 187 flares detected in \emph{Kepler} data from 1,547 solar-type stars, measured flare oscillation timescales of 10$^{2}$ to 10$^{3}$ s that were generally comparable to solar timescales.  In their review of solar and stellar QPPs, \citet{doo16} noted that the flare lengths attributed to magnetohydrodynamic (MHD) phenomenon found on other stars varied from $L=2.00\times 10^{10}$ cm to $4.00\times 10^{10}$ cm.  These length scales agree very well with those measured by other methods.  For instance, \citet{ben98} directly measured magnetic loop lengths of $L\sim2.2\times 10^{10}$ cm for the dM5.5 UV Ceti using the Very Long Baseline Array (VLBA).  Based on \emph{Extreme-Ultraviolet Explorer (EUVE)} photometry, \citet{haw95} determined coronal loop lengths of $L<(1.5-3.8)\times 10^{10}$ cm for the dM3.5e flare star AD Leonis, and $L=2.6\times 10^{10}$ cm for dM1 AU Mic.  Thus, our derived estimates of the HD 189733A flare lengths are entirely compatible with those measured on other stars, as summarized in Table 3, and do not lend support to the hypothesis of exoplanet-induced stellar activity.

Alternatively, the periodicity observed in the HD 189733A flares may not be the result of magnetoacoustic waves.  Instead of oscillations in a single flare, the multiple peaks observed in the X-ray light curve may point to the superposition of unassociated flares on the one hand, or ``sympathetic flaring'' on the other.  Thus far, stellar researchers have analyzed complex, multipeak flaring events from two different perspectives: flares may erupt near the same time by chance anywhere on the stellar surface and be causally disconnected, or a flare in one region may trigger activity in nearby active regions.  These two perspectives have existed in the solar literature since 1936, but only recently has an additional perspective been added: that causally connected, sympathetic flaring may be triggered in regions a full solar diameter apart, as occurred, for example, during the 2010 August 1-2 series of flares and CMEs observed via \emph{Solar Dynamics Observatory (SDO)} and \emph{Solar TErrestrial RElations Observatory (STEREO)} \citep{sch11}.  Although statistical arguments of increasing complexity have been advanced to bolster these perspectives, clearly these investigations are insufficient to get at the true nature of the physical mechanisms involved, as can be accomplished with multiwavelength observational analysis such as this one.  

Regardless of the exact mechanism involved, we can examine the likelihood of witnessing complex flares on HD 189733A.  \citet{dav14} examined the rate of complex flaring among 3,737 \emph{Kepler} white-light flare events with durations $\geq$10 min, presumably from a variety of planet-hosting systems.  They categorized these flares both by eye, and via an automated method that decomposed the complex flares into 1-10 individual flares based on the Bayesian information criterion.  They found that 65-95\% of the white flares in their sample with durations similar to the X-ray flares observed by \citet{pil10,pil11,pil14} (9-11 ks) were complex.  This demonstrates that the flare flickering timescales observed at HD 189733A, and the length scales they imply, are not unusual for long duration stellar flares, and probably do not point to an origin in ``star-planet interactions.''

\section{Conclusion}
In this paper, we have continued our investigations of potential ``star-planet interactions'' within the HD 189733 system.  We determined the locations of photometric maxima within APT, \emph{MOST}, and Wise data in search of bright regions associated with shocks and chromospheric and photospheric hotspots that may be caused by steady accretion from the exosphere of the exoplanet HD 189733b as \citet{pil10,pil11,pil14,pil15} alleged.  Based on our Kolmogorov-Smirnov and Lomb-Scargle periodogram analyses, we find no evidence for persistent bright regions that have locations synchronized to the exoplanet orbital period.  While bright regions may persist for a few rotational periods, this behavior is entirely consistent with the normal evolution of active regions and plage on stars.  Our results are consistent with and build off the dark region photometric and radio analyses presented in ROME I.  Our results are also consistent with ZDI measurements of the photospheric magnetic topology of HD 189733A which find no evidence for ``star-planet interactions'' and find that the surface field evolves significantly without stable features across epochs \citep{far10,far17}.

We have compiled multiwavelength observations of the HD 189733 system and compared its properties with those observed in actively accreting star systems, such as CTTSs.  These findings are summarized below.
\begin{enumerate}
	\item The magnitude of the photometric variability observed at HD 189733A is tiny compared to that found in CTTS systems (1-2\% vs. 3-137\%).  However, CTTS photometric variability is correlated with the rotation period of the pre-main-sequence star, as is the photometric variability of HD 189733A.
	\item The radio and X-ray emissions from CTTS systems are correlated as described by the G\"{u}del-Benz relationship.  Our Arecibo Observatory data is not sensitive enough to detect such gyrosynchrotron emission; however, our survey would have been sensitive to coherent radio emissions such as that found at the CTTS T Tau Bb, if present.
	\item With a maximum accretion rate of $\dot{M}\sim1\times 10^{11}$ g s$^{-1}$, HD 189733A would accrete at least two orders of magnitude less gas than CTTSs, if it accretes at all.  This tiny accretion rate provides a natural explanation for why if HD 189733A accretes from its exoplanet companion, related emissions would be undetectably weak.
	\item The velocity centroids of the spectral lines \ion{C}{2}, \ion{N}{5}, \ion{Si}{2}, \ion{Si}{3}, and \ion{Si}{4} are blue \emph{and} redshifted $\Delta v\lesssim +$20 km s$^{-1}$.  In CTTS and WTTS systems, accreted lines are redshifted by $\Delta v\gtrsim +$10$\pm$5 km s$^{-1}$. 
	\item The FUV lines measured at HD 189733A have FWHM$\sim$40km s$^{-1}$, less than half those found at CTTSs (FWHM$\gtrsim$100 km s$^{-1}$); this indicates that the emitting gas does not experience a large velocity gradient, as is found in T Tauri systems when gas is accreted from the inner truncation radius of the circumstellar disk to the stellar photosphere.
	\item One discriminator of the type of T Tauri system is H$\alpha$ equivalent width, $W_{eq}$(H$\alpha$).  At HD 189733A, $W_{eq}$(H$\alpha$)$\sim$10$^{-2}$\AA, which is similar to WTTS systems, whereas much larger values, $W_{eq}$(H$\alpha$)$\sim$30 to 200$\AA$ are measured in CTTS systems.
	\item An accretion shock at HD 189733A would be undetectable, yet with a temperature $T\sim4\times 10^{6}$ K, such a shock would contribute to a soft X-ray excess as they do in CTTS systems.
\end{enumerate}

An alternative scenario to the ``star-planet interactions'' alleged to occur in the HD 189733A/b system is that instead intrinsic stellar activity has been mistaken for these interactions.  In ROME I, we advanced this view based on a compilation of multiwavelength data that demonstrated that flaring activity is readily observed at all electromagnetic wavelengths independent of exoplanet orbital phase.  Again, we find here that the various measurements obtained by \citet{pil10,pil11,pil14,pil15} are exactly as expected for intrinsic stellar activity.  In Section 3, we found that photospheric bright regions appear irrespective of exoplanet orbital phase, as expected by a model of intrinsic activity, and in contradiction to the hypothesis of bright regions associated with the ``knee'' and photospheric impact hotspot features.  We derived an exospheric evaporation rate, and hence, maximum accretion rate of $\dot{M}=1.58\times 10^{11}$ g s$^{-1}$ (Section 4.3) and from blue and redshifted FUV spectral lines, projected plasma motions of $\Delta v\sim\pm$20 km s$^{-1}$(Section 4.4).  This compares favorably with the upflow and downflow velocities in solar flare plasma \citep{fle11} and the ejected mass and plasma velocities measured in CMEs ($\dot{M}\sim 10^{14}$-$4\times 10^{16}$ g and $v\sim$20 to 3,500 km s$^{-1}$, respectively) \citep{pri14}.  We derived a maximum accretion luminosity of $L_{S}=2.88\times 10^{26}$ erg s$^{-1}$ (Section 4.5.2), while \citet{pil14} computed X-ray flare luminosities of $L_{X}\sim 10^{28}$ erg s$^{-1}$.  By comparison, CMEs typically release $\sim 10^{32}$ erg in both kinetic energy and heating/radiation.  We also note that even a modest flare can reach radiative luminosities of $L\sim 5\times 10^{26}$ erg s$^{-1}$ \citep{pri14}.  We computed flaring electron density and temperatures of $n_{e}\sim$10$^{10}$-10$^{11}$ cm$^{-3}$ and $T\sim$10$^{6-7}$ K which are similar to values measured on the Sun and other magnetically active stars such as active K dwarfs \citep{gud09}.  Finally, we computed temporal and plasma flare properties that yielded lengths ranging from $L\sim(3-10)\times 10^{10}$ cm, which are not atypical for the Sun or other stars.  Given these considerations, a far simpler explanation for the witnessed activity is that the primary star is merely intrinsically active.

Comparison of HD 189733A with CTTSs provides some avenues of exploration that would aid the resolution of the controversy surrounding the nature of the HD 189733 system.  One testable hypothesis is the detection and measurement of veiling of photospheric lines at particular rotational and/or orbital cycles, as detected, for instance, at TW Hya \citep{don11a,joh13}.  Another hypothesis to test is whether \ion{He}{1} and \ion{Ca}{2} infrared triplet (IRT) lines correlate with each other across rotational and/or orbital phase, as they are strong indicators of accretion processes in CTTSs (e.g., TW Hya; \citet{don11b}).  A third test of the similarity between HD 189733A and T Tauri systems would be the measurement and analysis of the \ion{C}{4} (1548.2, 1550.8 \AA) and \ion{He}{2} (1670.5\AA) lines as compared to CTTSs and WTTSs (e.g., \citet{ard13}).  Future work should also aim to better measure the exospheric evaporation rate from HD 189733b and characterize its variability, which will inform estimates of the maximum accretion rate and emission luminosity from any shock region.  Moreover, despite the excitement that the notion of ``star-planet interactions'' has generated within the scientific community, it is important that any hypothesis fits all of the data with as few assumptions as possible.

\section{Acknowledgments}

We thank the referee for insightful comments that have improved the clarity of the text and presentation of the data analysis within this manuscript.  L.W.L. acknowledges support from NSF AST-1910364.  This research has made use of NASA's Astrophysics Data System.  The Lomb-Scargle periodogram analysis presented here is derived from source code made available by Brett Shoelson via MathWorks File Exchange.

\facility{Arecibo.}
\software{MATLAB.}

\clearpage
\begin{deluxetable}{lcccccc}
\tabletypesize{\scriptsize}
\tablecolumns{7}
\tablewidth{0pt}
\tablecaption{Photometric Bright Region Measurements and Phases}
\tablehead{
	\colhead{Observation}&
	\colhead{Observation}&
	\colhead{Bright Region}&
	\colhead{Bright Region}&
	\colhead{Bright Region}&
	\colhead{Source}&
	\colhead{Time Series}\\
	\colhead{Start (JD)}&
	\colhead{Finish (JD)}&
	\colhead{Midpoint (JD)}&
	\colhead{Rotation Cycle}&
	\colhead{Orbit Cycle}&
	\colhead{}&
	\colhead{Reference}	
}
\startdata
2453652.00 & 2453663.00 & 2453656.71 & 2.28 & 12.31 & APT & 1\\
2453663.00 & 2453672.00 & 2453669.02 & 3.30 & 17.86 & APT & 1\\
2453672.00 & 2453678.00 & 2453676.08 & 3.89 & 21.04 & APT & 1\\
2453678.00 & 2453686.00 & 2453680.74 & 4.28 & 23.15 & APT & 1\\
2453686.00 & 2453692.00 & 2453688.87 & 4.96 & 26.81 & APT & 1\\
2453692.00 & 2453700.00 & 2453694.13 & 5.40 & 29.18 & APT & 1\\
2453700.00 & 2453710.00 & 2453705.85 & 6.37 & 34.46 & APT & 1\\
2453810.00 & 2453820.00 & 2453817.22 & 15.65 & 84.66 & APT & 1\\
2453837.00 & 2453845.00 & 2453841.85 & 17.70 & 95.76 & APT & 1\\
2453845.00 & 2453851.00 & 2453847.11 & 18.14 & 98.13 & APT & 1\\
2453851.00 & 2453860.00 & 2453855.00 & 18.80 & 101.69 & APT & 1\\
2453860.00 & 2453870.00 & 2453867.91 & 19.88 & 107.51 & APT & 1\\
2453875.00 & 2453887.00 & 2453882.02 & 21.05 & 113.87 & APT & 1\\
2453887.00 & 2453902.00 & 2453897.09 & 22.31 & 120.66 & APT & 1\\
2453902.00 & 2453910.00 & 2453907.85 & 23.20 & 125.51 & APT & 1\\
2453949.00 & 2453962.00 & 2453960.60 & 27.60 & 149.29 & \emph{MOST} & 2\\
2453995.00 & 2454010.00 & 2454003.79 & 31.20 & 168.76 & APT & 1\\
2454010.00 & 2454022.00 & 2454015.75 & 32.20 & 174.15 & APT & 1\\
2454022.00 & 2454032.00 & 2454024.96 & 32.96 & 178.30 & APT & 1\\
2454032.00 & 2454045.00 & 2454039.54 & 34.18 & 184.87 & APT & 1\\
2454045.00 & 2454058.00 & 2454050.78 & 35.12 & 189.94 & APT & 1\\
2454058.00 & 2454070.00 & 2454062.98 & 36.13 & 195.44 & APT & 1\\
2454070.00 & 2454082.00 & 2454075.89 & 37.21 & 201.26 & APT & 1\\
2454170.00 & 2454180.00 & 2454174.83 & 45.45 & 245.85 & APT & 1\\
2454190.00 & 2454200.00 & 2454195.27 & 47.16 & 255.07 & APT & 1\\
2454200.00 & 2454210.00 & 2454206.27 & 48.07 & 260.02 & APT & 1\\
2454210.00 & 2454225.00 & 2454218.11 & 49.06 & 265.36 & APT & 1\\
2454225.00 & 2454236.00 & 2454232.22 & 50.24 & 271.72 & APT & 1\\
2454236.00 & 2454247.00 & 2454243.93 & 51.21 & 277.00 & APT & 1\\
2454247.00 & 2454258.00 & 2454256.37 & 52.25 & 282.61 & APT & 1\\
2454258.00 & 2454270.00 & 2454262.23 & 52.74 & 285.25 & APT & 1\\
2454270.00 & 2454283.00 & 2454276.10 & 53.89 & 291.50 & APT & 1\\
2454299.00 & 2454311.00 & 2454309.34 & 56.66 & 306.48 & \emph{MOST} & 3\\
2454311.00 & 2454321.00 & 2454319.78 & 57.53 & 311.19 & \emph{MOST} & 3\\
2454321.00 & 2454325.00 & 2454324.06 & 57.89 & 313.12 & \emph{MOST} & 3\\
2454325.00 & 2454329.80 & 2454329.13 & 58.31 & 315.40 & \emph{MOST} & 3\\
2455130.00 & 2455136.00 & 2455135.63 & 125.52 & 678.92 & APT and Wise & 4\\
2455136.00 & 2455141.00 & 2455140.00 & 125.88 & 680.89 & APT and Wise & 4\\
2455141.00 & 2455148.00 & 2455147.00 & 126.47 & 684.05 & APT and Wise & 4\\
2455148.00 & 2455152.00 & 2455151.50 & 126.84 & 686.08 & APT and Wise & 4\\
2455152.00 & 2455158.00 & 2455157.75 & 127.36 & 688.89 & APT and Wise & 4\\
2455158.00 & 2455164.00 & 2455163.13 & 127.81 & 691.32 & APT and Wise & 4\\
2455164.00 & 2455171.25 & 2455168.75 & 128.28 & 693.85 & APT and Wise & 4\\
2455321.43 & 2455328.00 & 2455324.92 & 141.29 & 764.24 & APT and Wise & 4\\
2455328.00 & 2455334.00 & 2455330.00 & 141.72 & 766.53 & APT and Wise & 4\\
2455334.00 & 2455341.00 & 2455336.88 & 142.29 & 769.63 & APT and Wise & 4\\
2455341.00 & 2455347.00 & 2455342.42 & 142.75 & 772.13 & APT and Wise & 4\\
2455347.00 & 2455350.00 & 2455348.38 & 143.25 & 774.82 & APT and Wise & 4\\
\enddata
\tablecomments{These measurements represent local maxima in the time series of the photometric flux from HD 189733A. See Section 2 for more details on the measurement of parameters and the computation of rotation and orbital cycles. Bright regions appear to be correlated in rotational, but not orbital, phase. {\bf Time series references:} (1) \citet{hen08}; (2) \citet{mill08}; (3) \citet{boi09}; (4) \citet{sing11}.  Reference 1 (3) time series are in HJD (BJD).}
\end{deluxetable}

\clearpage
\begin{deluxetable}{lcccccc}
\tabletypesize{\footnotesize}
\tablecolumns{7}
\tablewidth{0pt}
\tablecaption{Flare Density and Temperature Characteristics}
\tablehead{
	\colhead{Diagnostic}&
	\colhead{Ratio(s)}&
	\colhead{$n_{e}$ flare}&
	\colhead{$n_{e}$ quiescence}&
	\colhead{$T_{flare}$}&
	\colhead{$T_{quiescence}$}&
	\colhead{References}\\
	\colhead{}&
	\colhead{}&
	\colhead{($\times 10^{10}$cm$^{-3}$)}&
	\colhead{($\times 10^{10}$cm$^{-3}$)}&
	\colhead{($\times 10^{6}$K)}&
	\colhead{($\times 10^{6}$K)}&
	\colhead{}
}
\startdata
\ion{O}{7} & $R$ & $<$16 & 3.2-13 & 10.4-11.5 & 2.1-2.8 & \citet{pil11}\\
\ion{Ne}{9} & $R$ & - & $<$11 & 10.4-11.5 & 2.1-2.8 & \citet{pil11}\\
\ion{O}{7} & $R$ & 3-10 & 3-10 & 9.5 & 2.2 & \citet{pil14}\\
\hline
\ion{O}{7} & $R,G$ & 1.1 & 8.9 & 2 & $<$2 & This work\tablenotemark{a}\\
\ion{Ne}{9} & $R,G$ & 120 & - & $<$2 & $<$2 & This work\tablenotemark{a}\\
\ion{O}{7} & $R$ & 0.1-1.0 & 2-10 & 2 & $<$2 & This work\tablenotemark{b}\\
\ion{Ne}{9} & $R$ & 50-100 & - & $<$2 & $<$2 & This work\tablenotemark{b}\\
\hline
Solar & - & 10 & 0.1 & 10-40 & 1-6.3 & \tablenotemark{c}\\
Magnetic Stars & - & 1-10 & 1-10 & 5-50 & $<$10 & \tablenotemark{c}\\
\enddata
\tablecomments{See Section 4.5.1 for discussion and details.}
\tablenotetext{a}{Densities from $R$ ratios \citep{gud09}; temperatures from $G$ ratios \citep{por01}.}
\tablenotetext{b}{Temperature and density constraints from \citet{por01}.}
\tablenotemark{c}{Densities from \citet{asc05}; temperatures from \citet{gud09}.}
\end{deluxetable}

\begin{deluxetable}{lccccc}
\tabletypesize{\footnotesize}
\tablecolumns{6}
\tablewidth{0pt}
\tablecaption{Comparison of Stellar Flare Lengths}
\tablehead{
	\colhead{Flare Type}&
	\colhead{Object}&
	\colhead{Length}&
	\colhead{Length}&
	\colhead{Length}&
	\colhead{References}\\
	\colhead{}&
	\colhead{}&
	\colhead{($\times 10^{10}$cm)}&
	\colhead{($R_{\ast}$)}&
	\colhead{(au)}&
	\colhead{}
}
\startdata
Standing Wave\tablenotemark{a} & HD 189733A & 40 & 4.0 & 0.007 & \citet{pil14}\\
Standing Wave\tablenotemark{b} & HD 189733A & 25 & 4.7 & 0.017 & This Work\\
Standing Wave\tablenotemark{c} & HD 189733A & 11 & 2.1 & 0.0074 & This Work\\
Pressure Balance & HD 189733A & 6.0 & 1.1 & 0.0040 & This Work\\
Radiative Cooling & HD 189733A & 3.0 & 0.57 & 0.0020 & This Work\\
\hline
Flares-Typical & Sun & 2.1 & 0.30 & 0.0014 & \citet{def07}\\
Flares- QPP\tablenotemark{d} & Sun & 5.00-14.1 & 0.72-2.01 & 0.0033-0.0094 & \citet{fou05,fou10}\\
Flares- QPP & Various & 2.00-4.00 & - & 0.00134-0.00268 & \citet{doo16}\\
VLBA Measurement & UV Ceti & 2.2 & 2.2  & 0.0015 & \citet{ben98}\\
Radiative Cooling\tablenotemark{e} & AD Leo & $<$3.8 & $<$1.3 & $<$0.0025 & \citet{haw95}\\
Radiative Cooling\tablenotemark{e} & AU Mic & 2.6 & 0.47 & 0.0017 & \citet{haw95}\\
\enddata
\tablecomments{See Section 4.6 for more details.}
\tablenotetext{a}{Values given are reported in the cited text.}
\tablenotetext{b}{Revised to account for standing wave physics, but still using $\tau=9000$ s oscillation period.}
\tablenotetext{c}{Same as (b), but using the most statistically significant oscillation period, $\tau=4000$ s.}
\tablenotetext{d}{Assuming QPPs are attributed to fast kink mode oscillations.}
\tablenotetext{e}{Based on \emph{Extreme Ultraviolet Explorer (EUVE)} photometry.}
\end{deluxetable}

\clearpage
\begin{figure}
\plotone{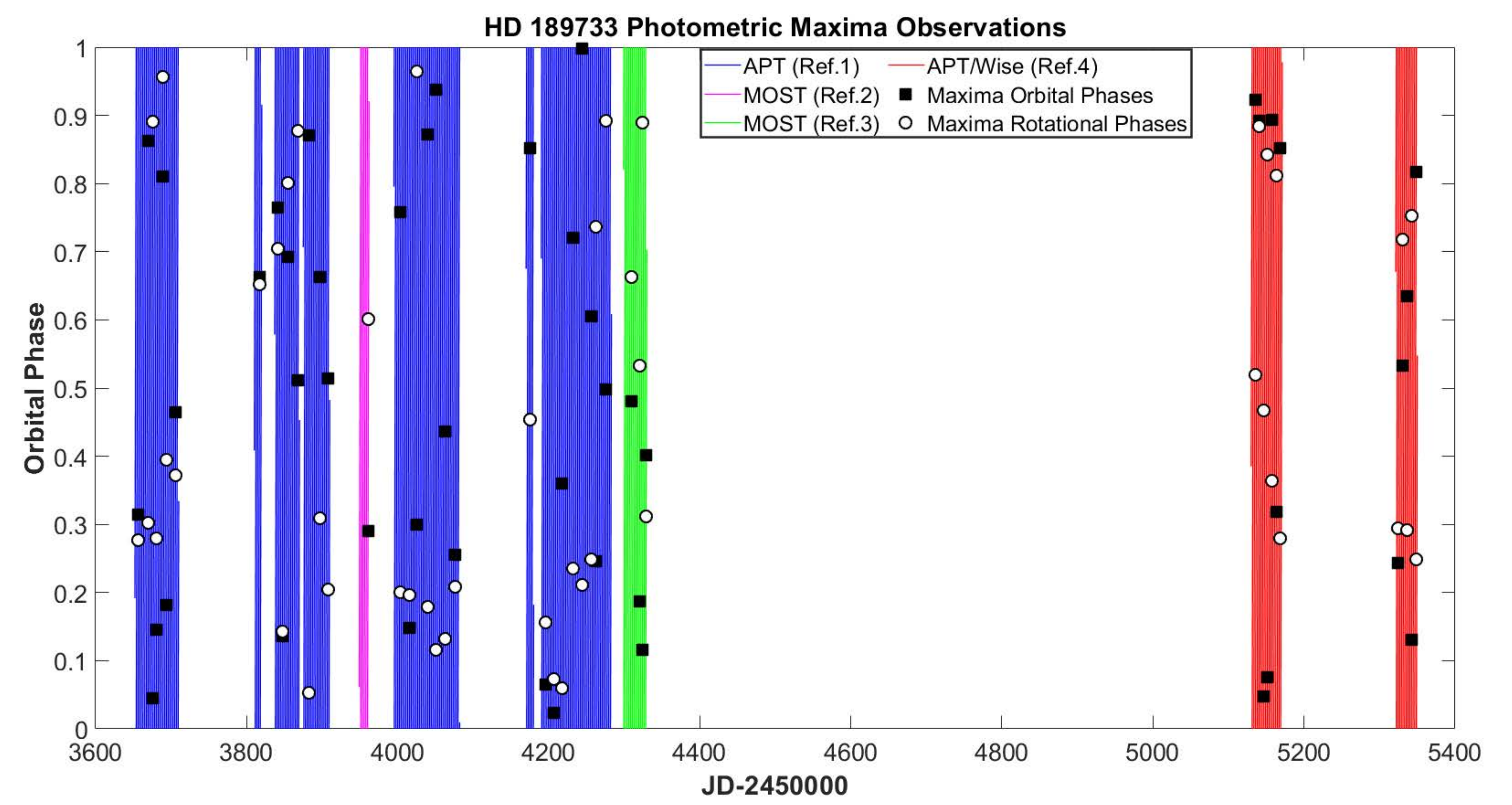}
\caption{Graphical representation of HD 189733A photometric maxima and observing epochs.  Blue, magenta, green, and red bands represent the dense observation of multiple HD 189733b orbital cycles in time, as recorded by APT \citep{hen08}, \emph{MOST} \citep{mill08}, \emph{MOST} \citep{boi09}, and APT and Wise \citep{sing11}, respectively.  Every cycle is traced out as a thin line within the band.  Black squares and white circles denote the orbital and rotational phases of the photometric maxima, respectively.  Note that circles are more tightly clustered across observing epochs, but squares are rarely clustered; this indicates that the bright regions are correlated across stellar rotational phase, but not exoplanet orbital phase.}
\end{figure}

\begin{figure}
\plotone{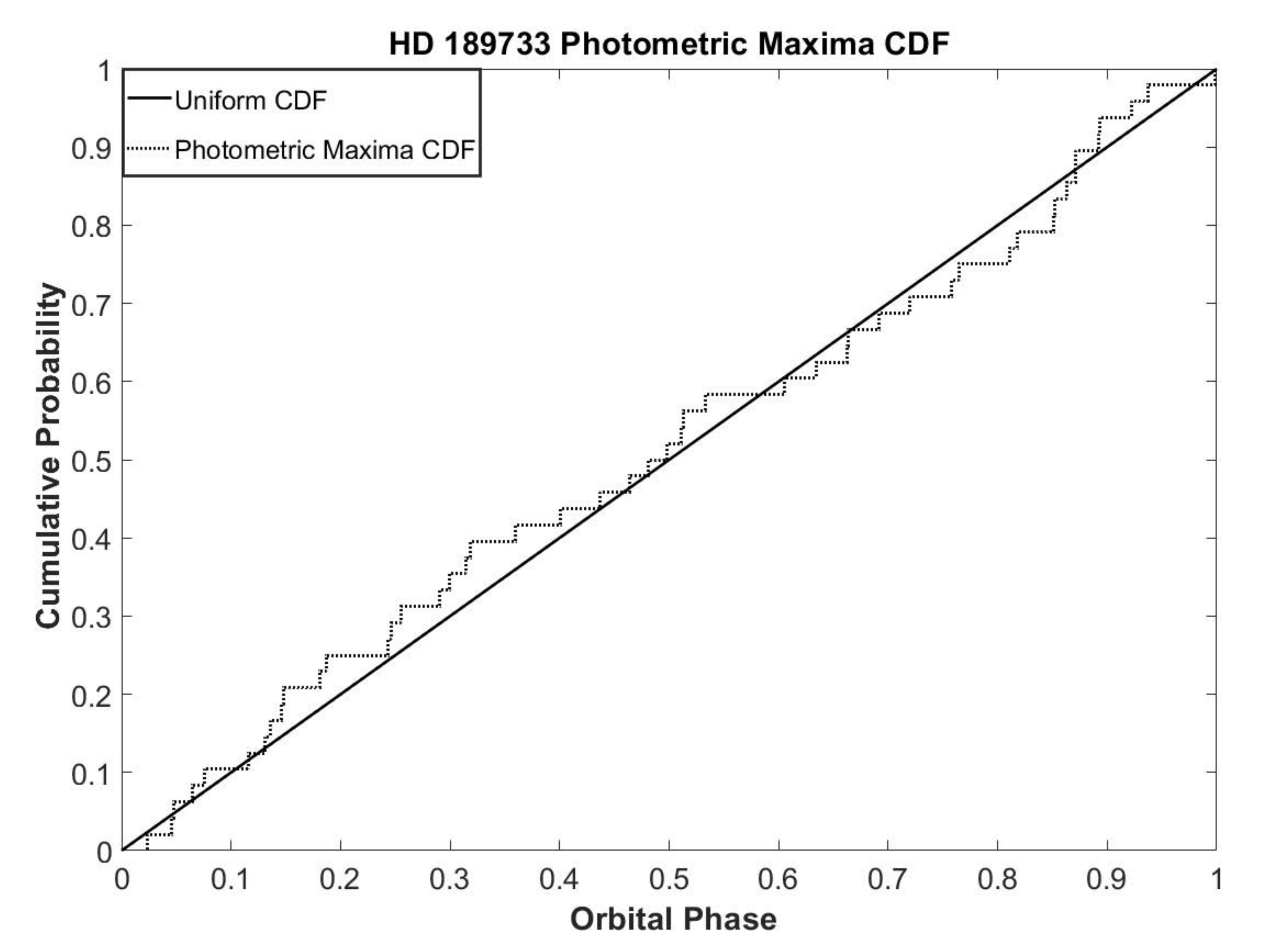}
\caption{Comparison of the empirical photometric maxima and modeled uniform cumulative distribution functions (CDFs) as a function of exoplanet orbital phase.  The deviation of the empirical CDF from the model is not statistically significant and indicates that they are drawn from the same underlying distribution.  Thus, bright regions are randomly distributed across the stellar surface independent of orbital phase.}	
\end{figure}

\begin{figure}
\plotone{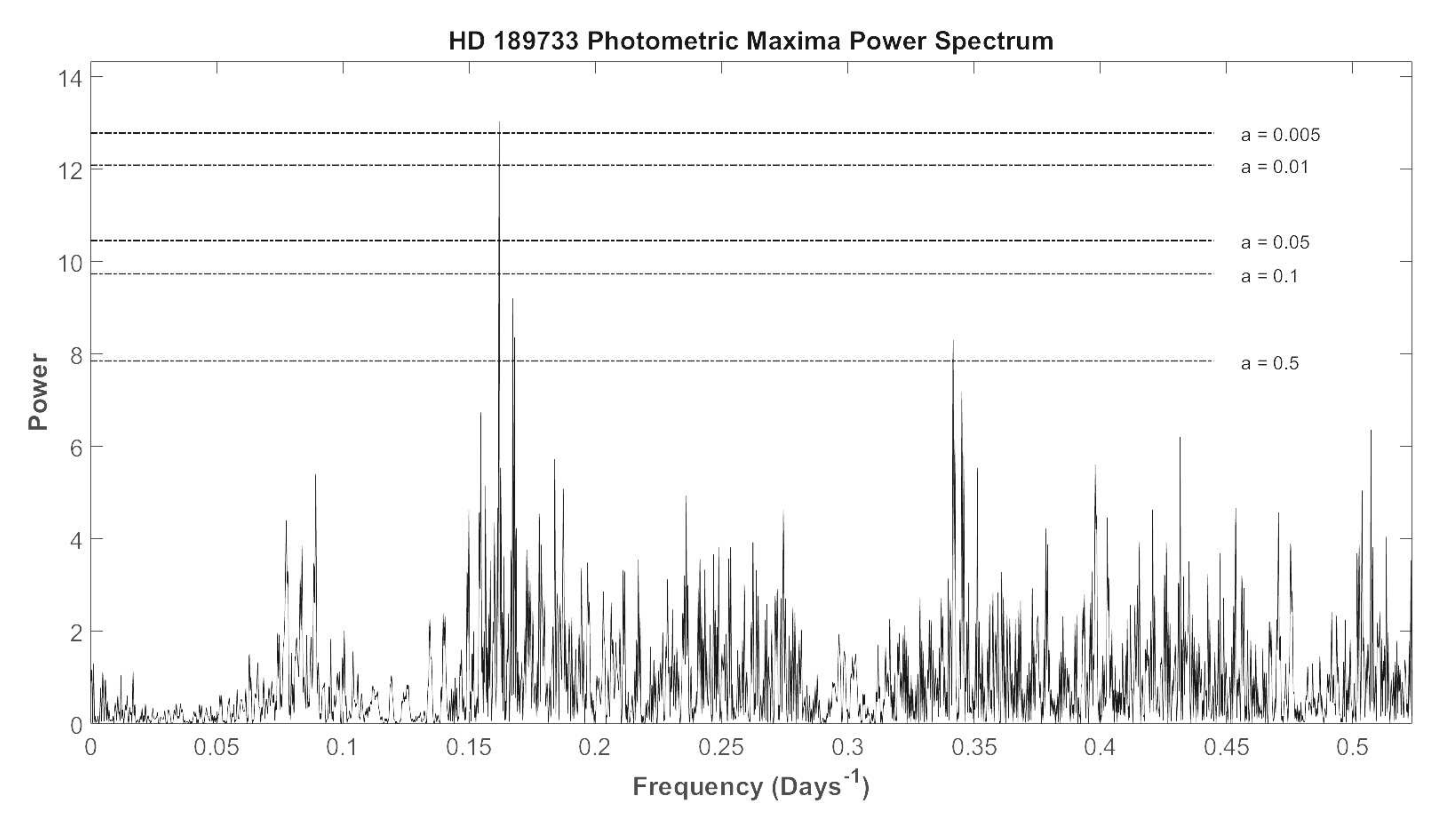}
\caption{Lomb-Scargle periodogram of HD 189733A photometric maxima listed in Table 1.  The only statistically significant frequency found is 0.16196478 day$^{-1}$ ($P$=6.1742 d).  Frequencies associated with the planetary orbital period (2.219 d, or $f=$0.451 d$^{-1}$) and the beat frequency between the stellar rotation period and the exoplanetary orbital period (2.5-2.7 d, or $f=$0.370-0.400 d$^{-1}$) are statistically insignificant.  Thus, this periodogram analysis does not lend support to the hypothesis that HD 189733A accretes from its exoplanet in any significant way that would generate a hot emitting gas region as \citet{pil15} proposed.  See Section 3.2 for more details.}
\end{figure}

\end{document}